\documentclass[aps,pre,groupedadress,nofootinbib,a4paper,showpacs,twocolumn]{revtex4-1}

\usepackage{paralist}		
\usepackage{color}		
\usepackage{times}
\usepackage{amsfonts}		
\usepackage{amsmath}		
\usepackage{amssymb}
\usepackage{latexsym}
\usepackage{graphicx}		
\usepackage{subfigure}
\usepackage{fullpage} 		
\usepackage{setspace}
\usepackage{hyperref}
\usepackage{textcomp}

\begin{document}

\title{Geometrical aspects of quantum walks on random two-dimensional structures}
\author{Anastasiia Anishchenko, Alexander Blumen, and Oliver Muelken}

\affiliation{
Physikalisches Institut, Universit\"at Freiburg,
Hermann-Herder-Stra{\ss}e 3, 79104 Freiburg, Germany}
 
\date{\today}

\begin{abstract}
We study the transport properties of continuous-time quantum walks (CTQW)
over finite two-dimensional structures with a given number of randomly
placed bonds and with different aspect ratios (AR). Here, we focus on the
transport from, say, the left side to the right side of the structure where absorbing
sites are placed. We do so by analyzing the long-time average of the
survival probability of CTQW. We compare the results to the
classical continuous-time random walk case (CTRW). For small AR
(landscape configurations) we observe only small differences between the
quantum and the classical transport properties, i.e., roughly the same
number of bonds is needed to facilitate the transport. However, with increasing AR
(portrait configurations) a much larger number of bonds is needed in the
CTQW case than in the CTRW case. While for CTRW the number of
bonds needed decreases when going from small AR to
large AR, for CTRW this number is large for small AR, has a minimum for the square
configuration, and increases again for increasing AR. We corroborate our
findings for large AR by showing that the corresponding quantum eigenstates are strongly localized
in situations in which the transport is facilitated in the CTRW case.

\end{abstract}

\maketitle

\section{Introduction}
Coherent dynamical processes in complex systems have become popular in different fields of science, ranging from chemistry and statistical physics \cite{muelken2011continuous, van1992stochastic} 
to quantum computation \cite{nielsen2010quantum}. The systems can be vastly different, say, optical waveguides \cite{lahini2011hanbury, heinrich2012disorder}, 
ultracold Rydberg gases \cite{deiglmayr2006coherent, westermann2006dynamics, mulken2007survival, cote2006quantum} or carbon nanotube networks \cite{kumar2005percolating, snow2003random, hu2004percolation}. Quantum mechanically as well as classically, 
transport in these systems takes place over different topologies which can vary from very ordered (regular) lattices to randomly build networks of interacting nodes. 
Then, an excitation is created at one or more of the nodes: the dynamics of the excitation is then described in the classical (diffusive) case by continuous-time random walks 
(CTRW) and in the quantum case by continuous-time quantum walks (CTQW) \cite{muelken2011continuous}.

In many cases one is interested in the transport {\it through} a network, i.e., an excitation 
is created somewhere in the network and can leave the network at a given set of nodes. The topological influence on the dynamics is 
then captured in the survival probability of the excitation to remain within the network. Here, we consider the example of a set of $N$ 
disconnected nodes arranged on two-dimensional lattices of different aspect ratios (AR) to which we randomly add a fixed number of bonds, $B$, between axially nearest-neighboring nodes. 
This resembles the random two-dimensional lattices of nanotubes whose conductivity properties have been studied experimentally \cite{kumar2005percolating, snow2003random, hu2004percolation}. 
There, the interest was in the conductivity from, say, 
the left side of the lattice to the right side.

In order to elucidate the transport properties of such networks, we calculate for each $B$ the long-time behavior (LTB) of the survival probabilities 
for CTQW and compare them to the ones for CTQW. We define $p_{0.5}^{QW}=B_{0.5}^{QW}/B_{\rm max}$, where $B_{0.5}^{QW}$ is 
that number of bonds, out of the total number $B_{\rm max}$, which is needed in order 
for the LTB of the CTQW survival probability to have reached (roughly) the value $0.5$. The corresponding CTRW 
probability is $p_{0.5}^{RW}$. Clearly, for the same AR, $p_{0.5}^{QW}$ and $p_{0.5}^{RW}$ can be vastly different, as the quantum-mechanical localization 
of eigenstates may lead to higher $p$-values for CTQW than for CTRW, see also Ref. \cite{leung2010coined} for a study of discrete-time quantum walks.

Before continuing with our analysis we mention the obvious connection to percolation theory \cite{stauffer1994introduction, sahimi1994applications}. While we focus on the survival 
probabilities and their decay due to existing connections from left to right, classical bond percolation focusses on the (first) appearance of such a connection. In our case, 
typically several of these connections are needed in order to reach the values $0.5$ for the LTB of both, CTQW and CTRW. We further focus on the time-independent 
case where bonds are permanent, i.e., 
they cannot be removed from the lattice once they are placed. In dynamical percolation, bonds might also be removed, see Ref. \cite{darazs2012time, kollar2012asymptotic}.

The paper is organized as follows: Section \MakeUppercase{\romannumeral 2} 
introduces the general concepts of CTRW and of CTQW. Furthermore, it discusses the trapping model and the different two-dimensional systems considered here.
Section \MakeUppercase{\romannumeral 3} displays our numerical results obtained for lattices with different AR for classical and for quantum 
mechanical transport. The paper ends in Section \MakeUppercase{\romannumeral 4} with our conclusions.  

\section{Transport over random structures}

\subsection{General considerations}

We start by considering both classical and quantum transport over two-dimensional structures consisting of
$N_x\times N_y = N$ nodes. We denote the position of a site by $j=(j_x,j_y)$, with $j_x=1,\dots,N_x$ and $j_y=1,\dots,N_y$, 
i.e. $j_x$ and $j_y$ are integers which label the lattice 
in the $x$- and the $y$-directions. Several of these nodes get connected by the $B$-bonds distributed over the structure. This 
procedure leads to a group of clusters of sites. The information
about these bonds is encoded in the $N\times N$ connectivity matrix $\bf
A$ (see, for instance, \cite{muelken2011continuous}). The non-diagonal elements of $\mathbf{A}$: pertaining to two sites are $-1$ if the 
sites are connected by one of the $B$-bonds and zero otherwise. The diagonal element of $\mathbf{A}$ corresponding to 
a particular site is $f$, where $f$ equals the number of $B$-bonds to which the particular site belongs. Now, it is non-negative definite, i.e. 
all its eigenvalues are positive or zero. When the structure contains no
disconnected parts, $\bf A$ has a single vanishing
eigenvalue \cite{biswas2000polymer}. In the following we describe 
the dynamics of purely coherent and of purely incoherent transport by
using the CTQW and the CTRW models,
respectively \cite{farhi1998quantum}. In both cases, the dynamics depends
very much on the topology of the structure, i.e., on $\mathbf A$. 
In a bra-ket notation, an excitation localized at node $j$ will be viewed as being in the
state $|j\rangle \equiv |j_x\rangle\
\otimes |j_y\rangle \equiv |j_x,j_y\rangle$. The states  $\{|j\rangle \}$ form an
orthonormal basis set.  Classically, the transport over unweighted and undirected
graphs can be described by CTRW with the transfer matrix
$\mathbf{T}=-\gamma\mathbf{A}$ \cite{van1992stochastic, farhi1998quantum,
muelken2011continuous}; here, for simplicity, we assume equal transition rates $\gamma=1$ for all
the nodes.

\subsection{CTQW and CTRW}

Quantum mechanically, the set of states $\{ |j\rangle \}$ spans the whole accessible
Hilbert space. The time evolution of an excitation starting at
node $|j\rangle$ can be described by the discrete Hamiltonian $\bf H$; Fahri and Guttmann assumed in \cite{farhi1998quantum} that
$\bf H=-\bf T$ which defines the CTQW corresponding to a CTRW with a given transfer matrix $\bf T$. 

The CTRW and the CTQW transition probabilities from the state
$|j\rangle$ at time $t=0$ to the state $|k\rangle$ at 
time $t$ read \cite{muelken2011continuous}:

\begin{eqnarray}
&& p_{k,j}(t)=\langle k|\exp{(-\mathbf{T}t)}|j\rangle 
\label{eq:trans_prob_cl}\\ 
\mbox{and} \qquad && \pi_{k,j}(t)=\arrowvert\langle\ k|\exp{(-i\mathbf{H}t)}|\ j\rangle \arrowvert^2,
\label{eq:trans_prob}
\end{eqnarray}
respectively, where we assume $\hbar=1$ in Eq.(\ref{eq:trans_prob}).

\subsection{The role of absorption}

An excitation does not necessarily stay forever in a particular system: it can
either decay or get absorbed at certain sites. Since we assume 
the lifetime of the excitation to be much longer than all the other relevant time
scales, we neglect the global radiative decay. However, there are specific
nodes where the excitation can get absorbed (trapped). We call these nodes traps 
and denote their set by $\mathcal{M}$. We also denote by $M$ the number of elements in $\mathcal{M}$ \cite{muelken2007localization}. 
The presence of traps
leads to the decay of the probability to find the
excitation in the system as a function of time \cite{muelken2011continuous}. 
For a trap-free structure we denote the transfer matrix and the Hamiltonian by $\mathbf {T}_0$ and by 
$\mathbf {H}_0$, respectively. We assume the trapping operator $\hat{\mathbf \Gamma}$ to be given by a sum over all
trap-nodes $|m\rangle=|m_x,m_y\rangle$ \cite{muelken2011continuous, muelken2010coherent}:

\begin{equation}
\label{eq:trapping matrix}
 \hat{\mathbf{\Gamma}}= \sum_{m\in\mathcal{M}} \Gamma_m
| m \rangle \langle m|.
\end{equation}

Then $\bf T$ and $\bf H$ can be written as $\mathbf{T}= \mathbf {T}_0-\mathbf{\Gamma}$ and $\mathbf{H}= \mathbf {H}_0-i\mathbf{\Gamma}$. 
In the CTRW case the transfer 
matrix stays real; then the transition probabilities can be calculated as:

\begin{equation}
p_{k, j}(t) = \sum_{n=1}^{N}e^{-\lambda_nt} \langle k |\phi_n \rangle \langle \phi_n |j
\rangle.
\label{SurvProbCTRW}
\end{equation}
In Eq.(\ref{SurvProbCTRW}) $\lambda_n$ are the (real) eigenvalues $\lambda_n$ and the $|\phi_n\rangle$ are the eigenstates of $\mathbf{T}$ . 

In the quantum mechanical case, $\mathbf {H}$ is non-hermitian and can have up to $N$ complex eigenvalues $E_n=\epsilon_n-i\gamma_n$, ($n=1,\dots,N$). 
Then the transition probabilities read:

\begin{equation}
 \pi_{k, j}(t) = \left |
\sum_{n=1}^{N}e^{-i\epsilon_nt}e^{-\gamma_nt} \langle k|\psi_n \rangle \langle \widetilde \psi_n | j
\rangle \right|^2,
\label{SurvProbCTQW}
\end{equation}
where $|\psi_n \rangle$ and $\langle \widetilde \psi_n |$ are the right
and the left eigenstates of $\mathbf{H}$, respectively. Obviously, 
the imaginary parts $\gamma_l$ of $E_l$ determine the temporal decay of $\pi_{k, j}(t)$.

\subsection{Structures with different aspect ratios}

\begin{figure}[h]
\centering
\includegraphics[width=\columnwidth]{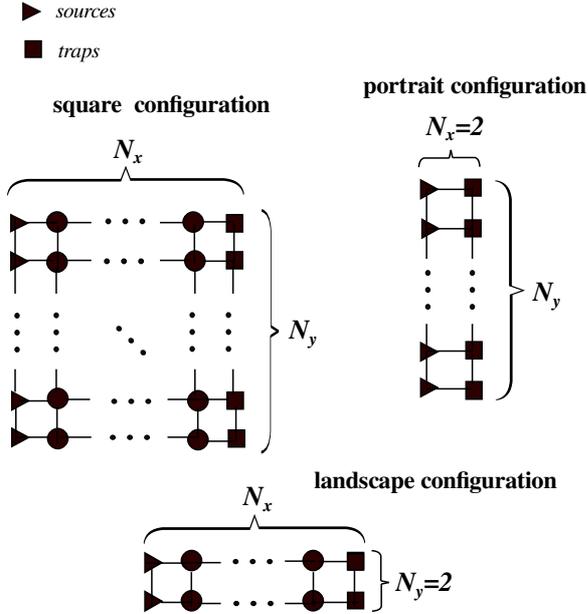}
\caption[networks]{Sketches of structures with square, portrait, and
landscape configurations. Here, the triangles denote possible sources and the squares denote the traps (sinks). 
The $B$-bonds are places on the horizontal and vertical connectivity segments.}
\label{fig:InCond}
\end{figure}

We now turn to specific examples two-dimensional structures with different AR, see Fig.~\ref{fig:InCond}. We distinguish the structures by their aspect ratio
$N_y/N_x$; in particular we denote the configurations of lattices with $N_y/N_x<1$ as ``landscapes'' 
and with $N_y/N_x>1$ as ``portraits''; the case $N_y/N_x=1$ is the square.

As stated above, we start from a set of $N=N_x\times N_y$ disconnected nodes, to which we randomly add $B$ bonds between nearest neighbor sites. This 
can be viewed as having bonds occupied 
with probability $p=B/B_{\rm max}$, with $B_{\rm max}$ being $B_{\rm max} = 2N_x N_y - (N_x+N_y)$. A simply connected component of
this graph is called a
cluster; every two nodes of such a cluster are connected to each other by at least one unbroken
chain of nearest-neighbors bonds.

We now focus on the transport in the $x$-direction. For this we depict the sites in the 
first column of the lattice by triangles and call them sources; their coordinates are $|1,l_y\rangle$, where $l_y=1,\dots,N_y$, see Fig.~\ref{fig:InCond}.
In a similar way, we depict the nodes of the last column by squares and call them traps (sinks). Their coordinates are $|N_x, m_y\rangle$, see
Fig.\ref{fig:InCond}. Thus, $\hat{\mathbf{\Gamma}}=
\sum_{m_y=1}^{N_y}\Gamma{(|N_x, m_y\rangle \langle m_y, N_x|)}$. Now, a typical process starts by exciting one of the sources. The process 
gets repeated by exciting another of the sources, and so forth. The classical and the quantum mechanical survival probabilities
$P(t)$ and $\Pi(t)$ are now:

\begin{equation}
\label{eq:mean survival probability clas}
P(t)= \frac{1}{N N_y}\sum_{l_y,k_y=1}^{N_y}
\sum_{k_x=1}^{N_x}\langle\
k_y,k_x|e^{-\mathbf{T}t}|1,l_y\rangle,
\end{equation}
and
\begin{equation}
\label{eq:mean survival probability}
\Pi(t)=
\frac{1}{NN_y}\sum_{l_y,k_y=1}^{N_y}
\sum_{k_x=1}^{N_x}\arrowvert\langle\
k_y,k_x|e^{-i\mathbf{H}t}|1,l_y\rangle \arrowvert^2.
\end{equation}
Note that in this way $p_{k,j}(t)$ and $\pi_{k,j}(t)$ are averaged over all possible
initial states $|1,l_y\rangle$ and over all possible final states $|k_x,k_y\rangle$.
Furthermore, the time evolution of $p_{k,j}(t)$ and $\pi_{k,j}(t)$  depends on the particular realization of the structure, since for a given, fixed $B$ the 
distribution of bonds and hence the structure is, in general, random. We evaluate interesting 
quantities through ensemble averaging over $R=1000$ random structure realisations and set:
\begin{equation}
\label{eq:ensemble}
 \langle...\rangle_R\equiv\frac{1}{R}\sum_{r=1}^{R}[...]_r.
\end{equation}
In such a way, we obtain 
ensemble-averaged survival probabilities $\langle P(t)\rangle_R$ and
$\langle \Pi(t)\rangle_R$ along with their long-time behavior (LTB)
$\langle P_\infty \rangle_R=\lim_{t \to \infty} \langle P(t)\rangle_R$ and
$\langle \Pi_\infty \rangle_R=\lim_{t \to \infty} \langle
\Pi(t)\rangle_R$. 

As stressed above, our interest is to determine for which values of $B$ $\langle P_\infty \rangle_R$ and $\langle \Pi_\infty \rangle_R$ 
reach the value $0.5$. We denote these values by $B_{0.5}^{(RW)}$ and $B_{0.5}^{QW)}$, respectively, and obtain thus
$p_{0.5}^{(RW)}=B_{0.5}^{(RW)}/B_{\rm max}$ and $p_{0.5}^{(QW)}=B_{0.5}^{(QW)}/B_{\rm max}$.

\section{Numerical results}

\subsection{$p_{0.5}^{(RW)}$ for CTRW and $p_{0.5}^{(QW)}$ for CTQW}

\begin{figure}[ht!]
\centering
\includegraphics[width=\columnwidth]{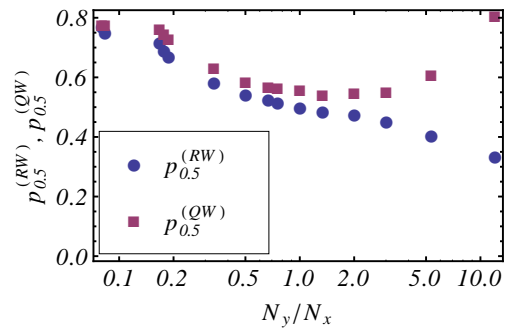}
\caption[networks]{Values of $p_{0.5}^{(RW)}$
and of $p_{0.5}^{(QW)}$ for different AR, $N_y/N_x$. Note
the logarithmic-linear scales.}
\label{fig:Pcr}
\end{figure}

Figure~\ref{fig:Pcr} summarises our findings for the classical $p_{0.5}^{(RW)}$ and for the quantum $p_{0.5}^{(QW)}$ as a function of the AR, namely of $N_y/N_x$.  
In general, we find $p_{0.5}^{(QW)}>p_{0.5}^{(RW)}$. For structures with $N_y/N_x<1$, i.e. in landscape
configurations,
$p_{0.5}^{(RW)}$ and $p_{0.5}^{(QW)}$ behave quite similarly as a function of $N_y/N_x$.
Now, increasing $N_y/N_x$ we find that $p_{0.5}^{(RW)}$ has a minimum at $N_y/N_x\approx1$, which is not the case for $p_{0.5}^{(QW)}$. 
For structures with $N_y/N_x>1$, i.e. in portrait configurations, the behavior of $p_{0.5}^{(RW)}$ and of $p_{0.5}^{(QW)}$ differs with increasing AR: In the CTRW case
$p_{0.5}^{(RW)}$ decreases with increasing
AR, reflecting the fact that the opposite ends get then closer, so that lower $p$-values are sufficient to ensure on efficient transport. 
In the CTQW case we find that for $N_y/N_x>1$ $p_{0.5}^{(QW)}$ increases with increasing AR, a quite counter-intuitive 
effect which we will discuss in detail in the following.

\begin{figure}[h]
\centering
\subfigure{\includegraphics[width=0.48\textwidth]{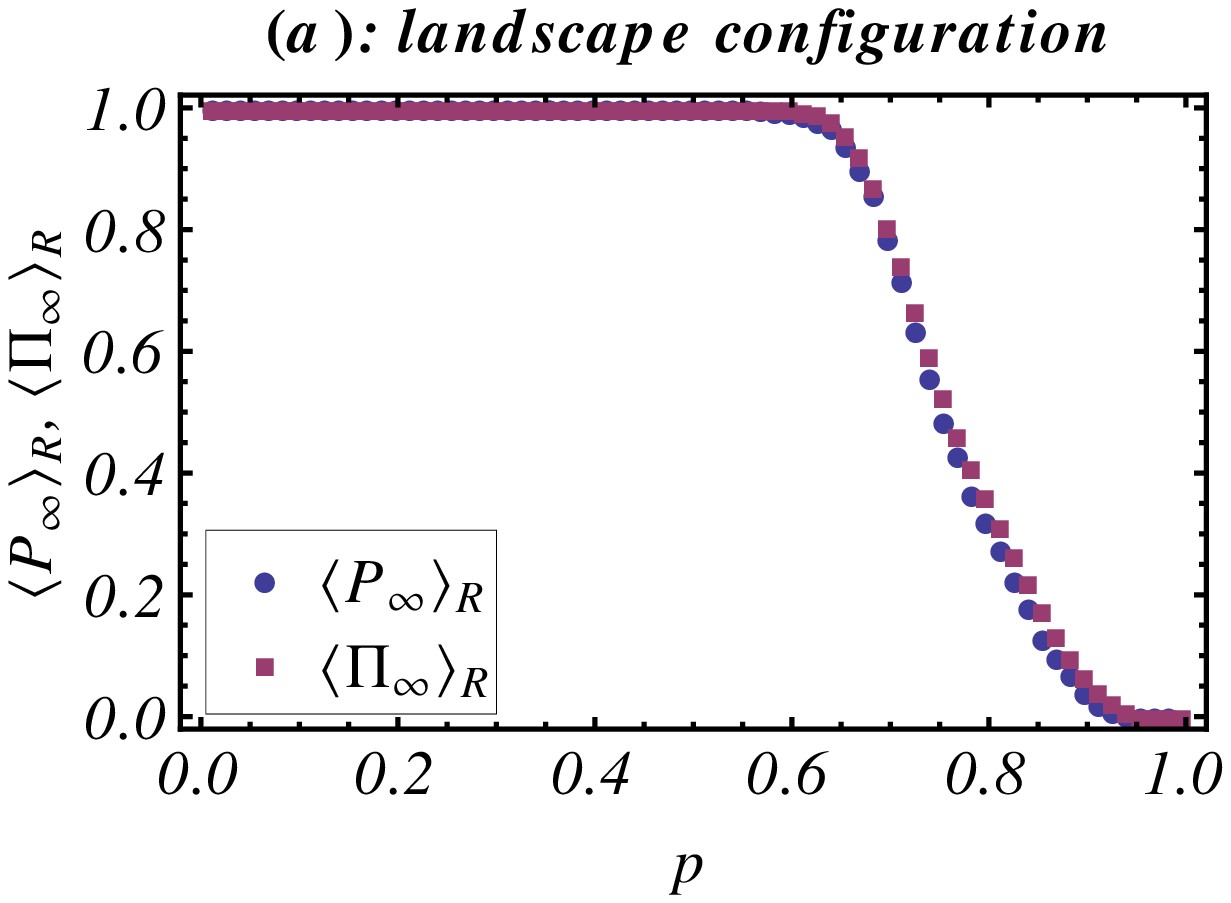}}
\subfigure{\includegraphics[width=0.48\textwidth]{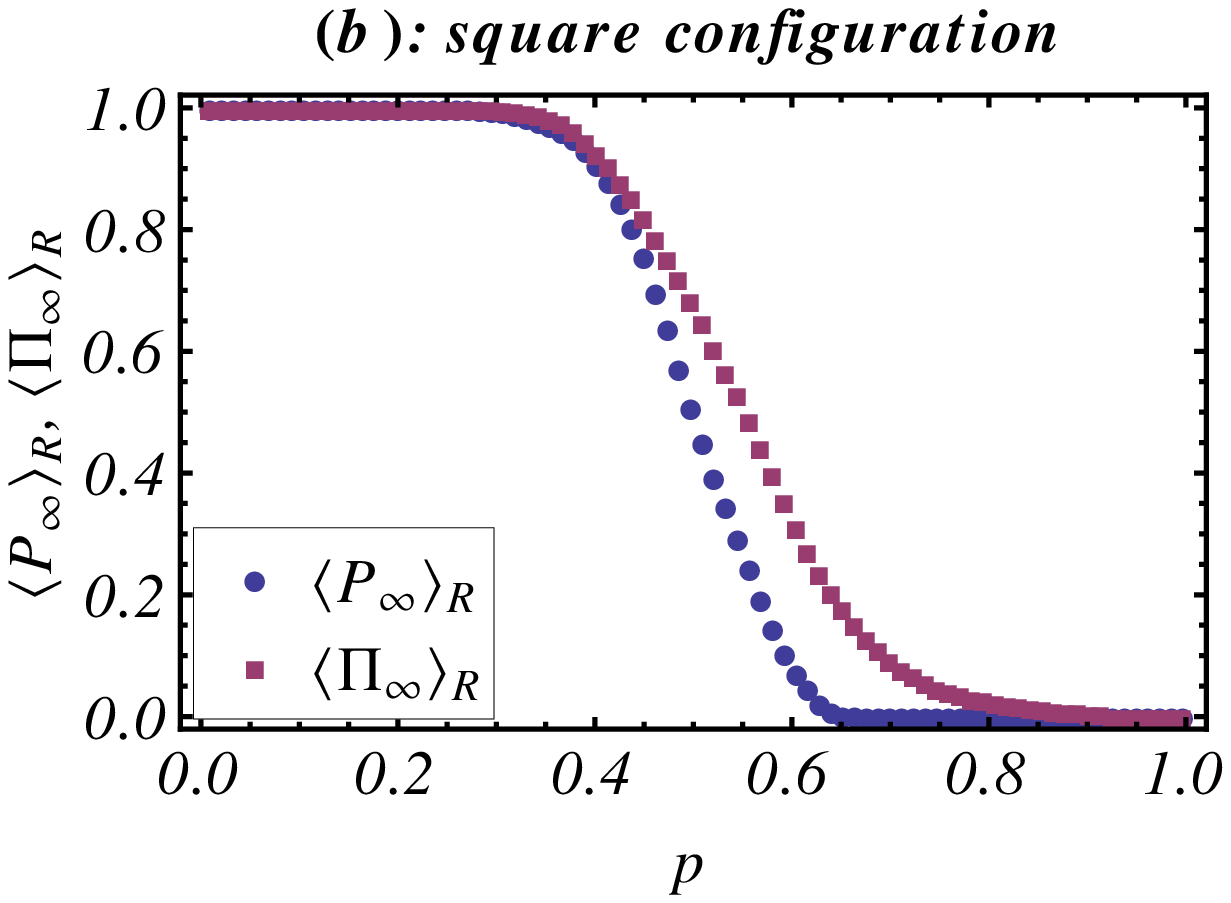}}
\subfigure{\includegraphics[width=0.49\textwidth]{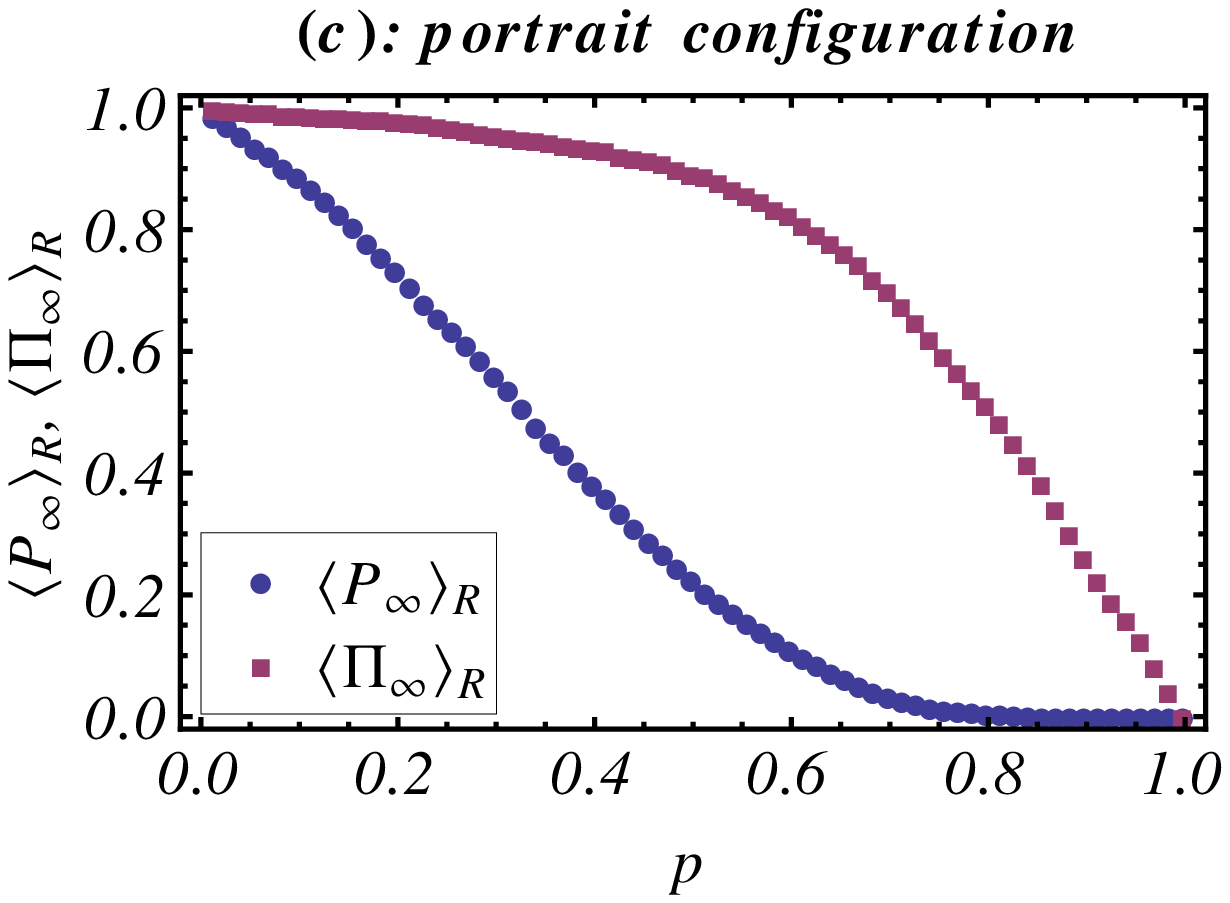}}
\caption[networks]{Values of $\langle P_\infty \rangle_R$ (circles) and of $\langle
\Pi_\infty \rangle_R$ (squares) as a function of $p$
for structures with different aspect ratios but with the same $B$ and roughly the
same $N$: (a) a landscape configuration with $24\times2$ nodes, (b) a square configuration with $7\times7$
nodes, and (c) a portrait configuration with $2\times24$ nodes.}
\label{fig:PercolationExamples}
\end{figure}

In Fig.~\ref{fig:PercolationExamples} we show particular examples of the
$p$-dependence of 
$\langle P_\infty \rangle_R$ and $\langle \Pi_\infty \rangle_R$
for structures with different AR but with roughly the
same total number $N$ of nodes. Displayed are:(a) a landscape configuration with $24\times2$ nodes, 
(b) a square configuration with $7\times7$
nodes, and (c) a portrait configuration with $2\times24$ nodes.
One observes as a function of $p$ the transition from states with very inhibited
transport, for which $\langle P_\infty \rangle_R$ and
 $\langle \Pi_\infty \rangle_R$ are very close to unity, to states in which the transport is very effective, 
 so that $\langle P_\infty \rangle_R$ and
 $\langle \Pi_\infty \rangle_R$ get very close to zero. From Fig.~\ref{fig:PercolationExamples} the values of $p_{0.5}^{(RW)}$ and of $p_{0.5}^{(QW)}$ 
 may be read off.
 Due to the finite size of the lattices
the transition region is rather broad; it gets sharper while increasing $N$. 
The difference in behavior between $\langle P_\infty
\rangle_R$ and $\langle \Pi_\infty \rangle_R$ is most evident for the portrait
configuration, see Fig.~\ref{fig:PercolationExamples}(c). Furthermore, in the portrait case the
CTRW $\langle P_\infty \rangle_R$ is smaller than in the square and in the landscape configurations. 
This is different than for the CTQW case, where $p_{0.5}^{(QW)}$ is larger than in the square 
and in the landscape configurations.

In the landscape configuration, the limit $N_y/N_x \to 0$ leads to
the situation of a very long (infinite) chain. In this case already one
broken bond is enough to inhibit transport, this is in line with our findings, both in the classical and in the quantum
mechanical cases, where we have $p_{0.5}^{(RW)} = p_{0.5}^{(QW)} = 1$. 

On the other hand, in the limit $N_y/N_x \to \infty$ one finds that for CTRW only a small number of bonds $B$, i.e., a small 
probability $p$ is sufficient to cause a drop in $\langle P_\infty \rangle_R$. This is readily seen in the limit $N_x=2$, when 
a horizontal bond is guaranteed in average when $B$ is around $3$ (one has for $N_x=2$ roughly twice as many vertical as horizontal bonds), i.e. 
for $p\simeq3/3N_y=1/N_y$. Such a bond connects a source to a trap and this $p$ value, $p\simeq1/N_y$ 
tends to zero as $N_y/N_x \to \infty$. 

The picture is not so simple in the CTQW case.  
Here, the survival
probability depends on specific features of the eigenstates
$|\psi_n\rangle$. If these are localized, transport from one node to the
other will be inhibited as in the Anderson localization
\cite{anderson1958absence}. In the next section we will analyze the eigenstates of
$\mathbf H$ in order to understand the relatively large values of $p_{0.5}^{(QW)}$ compared to $p_{0.5}^{(RW)}$ for
lattices with portrait configurations.

\subsection{Participation ratio and eigenstates}

We recall that the participation ratio $| \langle j |
\psi_{n,r}^{(0)} \rangle|^4$, where $|\psi_{n,r}^{(0)}\rangle$ is the
$n$th eigenstate of the $r$th realization of the $\mathbf{H}_0$, is a
measure of the localization of the different eigenstates. In
order to take the ensemble averaging into account, we introduce
\begin{equation}
\langle\Xi_{j,n}\rangle_R= \frac{1}{R}\sum_{r} | \langle j |
\psi_{n,r}^{(0)} \rangle|^4
\end{equation} 
as the ensemble averaged participation ratio
\cite{muelken2007small}.

\begin{figure*}[ht!]
\centering
\includegraphics[width=1.0\textwidth]{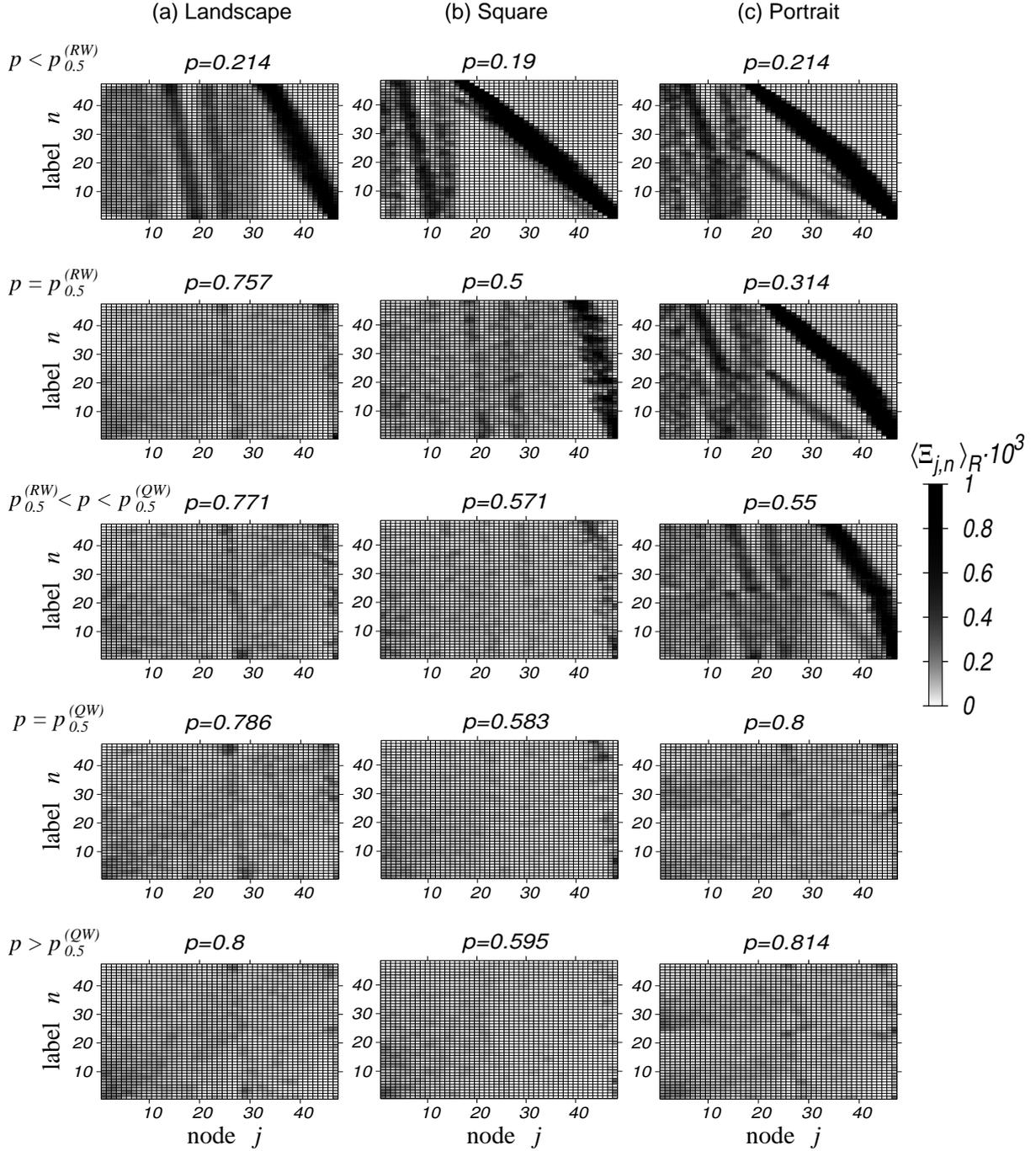}
\caption[part]{Ensemble averaged participation ratios
$\langle\Xi_{j,n}\rangle_R$ for different values of $p$, namely
$p<p_{0.5}^{(RW)}$, $p=p_{0.5}^{(RW)}$, $p_{0.5}^{(RW)} < p < p_{0.5}^{(QW)}$, $p=p_{0.5}^{(QW)}$, and $p>p_{0.5}^{(QW)}$,
for : 
(a) Landscape configuration for a lattice of $24\times2$ nodes. 
(b) Square configuration for a lattice of $7\times7$ nodes. 
(c) Portrait configuration for a lattice of $2\times24$ nodes. 
}
\label{fig:partratio}
\end{figure*}

Figure \ref{fig:partratio} shows in contour plots $\langle\Xi_{j,n}\rangle_R$ for lattices
whose configuration is (a) landscape, (b) square, and (c) portrait. Here, in each separate panel each row
reflects the average contribution of every node $|j\rangle$ of the lattice
to a given eigenstate $|\psi_{n,r}\rangle$. In order to see the transition
from the situation for $p<p_{0.5}^{(RW)}$ to the one for $p>p_{0.5}^{(QW)}$,
we present $\langle\Xi_{j,n}\rangle_R$ for distinct $p$ values, namely
for $p<p_{0.5}^{(RW)}$, for $p=p_{0.5}^{(RW)}$, for $p_{0.5}^{(RW)}<p<p_{0.5}^{(QW)}$, for $p=p_{0.5}^{(QW)}$, and for 
$p>p_{0.5}^{(QW)}$. Bright shadings correspond to low while dark shadings correspond to high
values of $\langle\Xi_{j,n}\rangle_R$. Therefore, localized dark regions
indicate localized eigenstates. These, in turn, will inhibit the transport.

This is well in line with the information obtained from Fig.~\ref{fig:PercolationExamples}, presented in Fig.~\ref{fig:PercolationExamples}(a) for the 
landscape configuration. We ramark that, as already noticeable from Fig.~\ref{fig:PercolationExamples}(a), for the landscape configuration the
quantum and the classical $p_{0.5}(t)$-probabilities lie very close together,
being $p_{0.5}^{(RW)}=0.757$ and $p_{0.5}^{(QW)}<0.786$. In the depicted case $p_{0.5}^{(RW)}$ and $p_{0.5}^{(QW)}$ differ only by $4\%$, i.e., for $N=48$
only by $2$ bonds in $B_{0.5}$. The eigenstates stray localized up to
$p=p_{0.5}^{(QW)}$, see the first panel in
Fig.~\ref{fig:partratio}(a). For $p>p_{0.5}^{(QW)}$ the eigenstates
get more delocalized, which is visible as the grey gets more evenly-distributed over the different nodes $n$. 

For the square configuration, Fig.~\ref{fig:partratio}(b), the relative difference between $p_{0.5}^{(RW)}$ and $p_{0.5}^{(QW)}$ is about twice as large as for the landscape configuration.
Here, one notices a strong localization of the eigenstates for $p$-values up to
$p_{0.5}^{(RW)}$, see the first two panels, while this effect is getting less pronounced for
larger values of $p$, this already indicates that quantum transport is
strongly inhibited for $p$-values below and close to $p_{0.5}^{(RW)}$.

This effect is even more enhanced for the portrait configuration, as may be seen from Fig.~\ref{fig:partratio}(c): Up to $p_{0.5}^{(RW)}$ one ramarks very 
strong localization. This persists even up to $p_{0.5}^{(QW)}=0.8$ which value is more than twice as large as
$p_{0.5}^{(RW)}=0.314$. In this particular example one has $N=48$, $B_{0.5}^{(RW)}=22$ and $B_{0.5}^{(RW)}=56$. This means that one needs 
more than twice more bonds in order to render the quantum transport as efficient as the classical one, in this particular portrait configuration. For smaller 
$B$ values, the eigenstates are too localized for the quantum transport to be efficient.

\section{Conclusions}

We have studied the coherent, continuous-time quantum transport on two-dimensional structures
of different aspect ratios $N_y/N_x$ with a given, fixed number $B$ of randomly
placed bonds. Having focused on three types of configurations -- landscape, square, and portrait -- we investigated the long-time probability for an excitation not to get trapped. 
Our analysis shows that
in the average the quantum excitation transport in the $x$-direction becomes very inefficient for structures 
with portrait configurations, i.e., for those where $N_y\gg N_x$. This is particularly remarkable, since the opposite holds for 
(incoherent) continuous-time random
walks, where the transport becomes more efficient when the $AR$ increases. This is rendered clear by our evaluations of the classical and
quantum mechanical probabilities $p_{0.5}^{(RW)}$ and
$p_{0.5}^{(QW)}$ which we have introduced in this article. 
The behavior in the quantum case can be understood based on an analysis of the corresponding eigenstates. Their participation ratios show that in portrait
configurations the eigenstates are still localized for probabilities $p$ such that 
$p_{0.5}^{(RW)} < p < p_{0.5}^{(RW)}$. Only for $p > p_{0.5}^{(RW)}$ the eigenstates do 
become delocalized and thus can readily support the transport. 

 \section*{Acknowledgments}
  
We thank Piet
Schijven for fruitful discussions.
Support from the Deutsche Forschungsgemeinschaft (DFG Grant No. MU2925/1-1), from  the Fonds
der Chemischen Industrie, from the Deutscher Akademischer Austauschdienst
(DAAD Grant No. 56266206), and from the Marie Curie International Research Staff Exchange 
Science Fellowship within the 7th European Community Framework Program SPIDER 
(Grant No. PIRSES-GA-2011-295302) is gratefully acknowledged.

%

\end{document}